\documentclass[a4paper,10pt]{article}
\usepackage[usenames,dvipsnames]{pstricks}
\usepackage[utf8]{inputenc}
\pdfoutput=1
\usepackage{lmodern} 
\usepackage[T1]{fontenc}
\usepackage{multicol}
\usepackage{epsfig}
\usepackage[fleqn]{amsmath}
\usepackage{amsmath,amsfonts}
\usepackage{textcomp}
\usepackage{graphicx}
\usepackage{subcaption}
\usepackage{authblk}
\usepackage[font={small,it}]{caption}
\usepackage{epstopdf}
\usepackage[pdfencoding=auto, psdextra]{hyperref}
\usepackage{bookmark}
\usepackage{float}
\usepackage[numbers]{natbib}
\hypersetup{pdfauthor={some author},pdftitle={eye-catching title}}
\usepackage {ifpdf}
\usepackage[margin=0.5 in]{geometry}

\title{\textbf{Quark-lepton complementarity model based predictions for $\theta_{23}^{PMNS}$ with neutrino mass hierarchy}}
\author{\small{Gazal Sharma\footnote{gazzal.sharma555@gmail.com}}}
\author{\small{Shankita Bhardwaj\footnote{shankita.bhardwaj982@gmail.com}}}
\author{\small{B. C. Chauhan\footnote{chauhan@associate.iucaa.in}}}
\author{\small{Surender Verma\footnote{s$\_$7verma@yahoo.co.in}}}
\affil{\textit{Department of Physics and Astronomical Science, }\\
\textit{ School of Physical and Material Sciences,}\\
\textit{ Central University of Himachal Pradesh (CUHP),
Dharamshala, Kangra (HP),
India 176215}}
\date{}

\begin{document}

\maketitle

\begin{abstract}
 After the successful investigation and confirmation of non zero $\theta_{13}^{PMNS}$ by various experiments, we are standing at a square where we still encounter a number of issues, which are to be settled. In this paper, we have extended our recent work towards a precise prediction of the 
 $\theta_{23}^{PMNS}$ mixing angle, taking into account the neutrino mass hierarchy. 
We parameterize the non-trivial correlation between quark (CKM) and lepton (PMNS) mixing 
matrices in quark-lepton complementarity (QLC) model as $V_{c}= U_{CKM}. \psi. U_{PMNS}$, where $\psi$ is a diagonal phase matrix. Monte Carlo simulations are used to estimate the texture of $V_{c}$ and compare the results with the standard Tri-Bi-Maximal (TBM) and Bi-Maximal(BM) structures of neutrino mixing matrix. We have predicted the value of $\theta_{23}^{PMNS} $ for normal and inverted neutrino mass hierarchies. The value of $\theta_{23}^{PMNS}$ obtained for two cases are about $1.3\sigma$ away from each other, implying the better precision can give us a strong hint for the type of neutrino mass hierarchy. 
\end{abstract}


\section{Introduction}
The recent results from various neutrino oscillation experiments \cite{1}  in past several years have provided us a very strong sign of neutrinos bieng massive, the mixing of lepton flavours and their oscilations. After the successful investigation of $\theta_{13}^{PMNS}$ \cite{2}, there are certain issues that are not settled yet, out of which one is the problem of mass hierarchy in the neutrino mass spectrum 
 and another is the quadrant of angle $\theta_{23}^{PMNS}$, are the challenges that are to be settled.

The phenomenon of quark and lepton flavor mixing is described by a $3\times3$ unitary matrix called Cabibbo-Kobayashi-Maskawa $(U_{CKM})$ and Pontecorvo-Maki-Nakagawa-Sakata $(U_{PMNS})$ respectively. After investigating the global data fits of various experimental results, so far we know the values for the $U_{PMNS}$ matrix which contains two large and a small mixing angles; i.e. the $\theta_{23}^{PMNS}$ $\approx$ $45^\circ$, the $\theta_{12}^{PMNS}\approx 34^\circ$ and the $\theta_{13}^{PMNS}\approx 9^{\circ}$. These results are observed along with the quark flavor mixing matrix $(U_{CKM})$, which is quite settled with three mixing angles that are small i.e. $\theta_{12}^{CKM}\approx 13^\circ$, 
$\theta_{23}^{CKM}\approx 2.4^\circ$ and $\theta_{13}^{CKM}\approx 0.2^\circ$, which clearly indicates about a disparity-cum-complementarity between quark and lepton mixing angles. Since, the quarks and leptons are fundamental constituents of matter and Standard Model(SM), the complementarity relation between these two families is seen as a consequence of a symmetry at some high energy scale. This complementarity termed as `Quark-Lepton Complementarity'(QLC) has been studied by various authors
\cite{3}.

The quark-lepton complementarity (QLC) relations hints about the depth of the structure that interrelates
quarks and leptons.
The disparity between the quark and lepton mixing angles has been expressed in
terms of the QLC relations, which can be written as
\begin{eqnarray}
\theta_{12}^{l}+\theta_{12}^{q}\simeq 45^\circ, \\ 
\theta_{23}^{l}+\theta_{23}^{q}\simeq 45^\circ, \\
\theta_{13}^{l}+\theta_{13}^{q}\simeq 0^\circ. 
\end{eqnarray}

The above QLC relations indicate that, on the basis of certain flavor symmetry there could be a quark-lepton symmetry at some different energy scale.

Possible consequences of QLC have been investigated in the literature and in particular
a simple correspondence between the $U_{PMNS}$ and $U_{CKM}$ matrices has been
proposed and analyzed in terms of a correlation matrix \cite{4}
\begin{equation}
V_{c}=U_{CKM} \cdot U_{PMNS},
\end{equation}
where $V_{c}$ is the correlation matrix defined as a product of $U_{PMNS}$ and $U_{CKM}$.
In section ({\bf\ref{sec:QLC}}), we discuss in brief the theory of the QLC model along with the investigation of correlation matrix ($V_{c}$) and the methodology that we have followed to obtain the desired results. 
According to the model procedure, after using the most credible texture of the correlation matrix
we derive the constraints on the $\theta_{23}^{PMNS}$ mixing angle for both normal and inverted neutrino mass hierarchies in the section~({\bf\ref{sec:reslt})}.
Finally, in section~({\bf\ref{sec:concl}}) we conclude and summarise our results.

\section{The QLC Model and Theoretical Framework }\label{sec:QLC}

 The texture of $V_{c}$ can be obtained under certain assumptions about the flavor structure of the
theory \cite{5}
\begin{equation}\label{eq:fund}
V_{c}=U_{CKM} \cdot \psi \cdot U_{PMNS},
\end{equation}

where $\psi$ is taken as diagonal matrix $\psi= diag(e^{(\iota \psi_i)})$ and the three phases $\psi_i$ are assumed to free parameters as they are not constrained by any of the current experimental evidences. 

 This is more convenient to do because in Grand Unified Theories (GUTs) \cite{6}, once quarks and leptons are kept in the same representation of the underlying gauge group, one has to include an arbitrary but non-trivial phases between the quark and
lepton mixing matrices in order to counter the
phase mismatch. We take

\[
U_{CKM} =
\begin{bmatrix}
1-\lambda^2/2-\lambda^4/8  & \lambda & A\lambda^3(1+ 
\lambda^2/2)(\bar{\rho}- \iota\bar{\eta})\\
-\lambda+A^2\lambda^5(1/2- \bar\rho-\iota\bar\eta) & 
1-\lambda^2/2-\lambda^4/8(1+4A^2) & A\lambda^2\\
A\lambda^3(1-\bar\rho-\iota\bar\eta) & 
-A\lambda^2+A\lambda^4(1/2-\bar\rho-\iota\bar\eta) & 1- 
A^2\lambda^4/2
\end{bmatrix}
+ {\cal O}(\lambda^6)
\] 
and
 \[U_{PMNS}=
 \begin{bmatrix}
  e^{\iota\phi_1}c_{12}c_{13} & 
e^{\iota\phi_2}c_{13}s_{12} & s_{13}e^{-\iota\phi}\\
  
e^{\iota\phi_1}(-c_{23}s_{12}-e^{\iota\phi}c_{12}s_{13}s_{
23}) & 
e^{\iota\phi_2}(c_{12}c_{23}-e^{\iota\phi}s_{12}s_{13}s_{
23}) & c_{13}s_{23}\\
  
e^{\iota\phi_1(-e^{\iota\phi_1}c_{12}c_{23}c_{13}+s_{12}s_
{23})} & 
e^{\iota\phi_2}(-e^{\iota\phi}c_{23}s_{12}s_{13}-c_{12}s_{
23}) & c_{13}c_{23}
 \end{bmatrix}.
\]

However, the values of quark($U_{CKM}$)\cite{1} and lepton($U_{PMNS}$)\cite{king} mixing parameters are at 1-$\sigma$ range

\minipage{0.45\textwidth}
\begin{eqnarray}\label{CKMnew}
 \lambda= 0.2255\pm0.0006,\\ \nonumber
  A=0.818\pm0.015,\\ \nonumber
 \bar\rho=0.124\pm0.024,\\ \nonumber
 \bar\eta=0.354\pm0.015,
\end{eqnarray}
\endminipage\hfill
\minipage{0.45\textwidth}
\begin{eqnarray}\label{PMNSa}
 \sin^{2}\theta_{13}=0.0218_{-0.0010}^{+0.0010},\\ 
\nonumber
 \sin^{2}\theta_{12}= 0.304_{-0.012}^{+0.013},\\ \nonumber
 \sin^{2}\theta_{23}=0.452_{-0.028}^{+0.052},\\ \nonumber
  \phi= (306^{\circ})_{-70}^{+39}.
\end{eqnarray}
\endminipage\hfill

For the unknown phases $\phi_1$, $\phi_2$ and the three $\psi_{i}$ ({\bf\ref{eq:fund}}), we vary their values between the open 
interval $[0,2\pi]$ in a flat distribution.\\
As per our model \cite{5} procedure, in order to constrain the value of $\theta_{23}^{PMNS}$ we 
use the inverse equation obtained
\begin{equation}
U_{PMNS}=(U_{CKM}\cdot \Psi)^{-1}\cdot V_{c}.
\end{equation}

We follow more familiar and democratic approach for the calculation of the correlation
matrix i.e. it may take any form of texture as suggested by theoretical
and experimental data from quark and lepton sectors. 
After doing Monte Carlo simulations we estimated the texture of the $V_c$ matrix.
We obtained predictions for $U_{PMNS}^{23}$ for the two cases of neutrino mass hierarchies i.e. normal hierarchy(NH) and inverted hierarchy(IH).

\section{Results}\label{sec:reslt}

The PMNS matrix obtained in case of normal hierarchy is

\minipage{0.45\textwidth}
\[
U_{PMNS}=
  \begin{bmatrix}
    0.68-0.87 & 0.40-0.67 &  0.02-0.35\\
    0.20-0.59 & 0.40-0.70 & 0.59-0.74 \\
    0.38-0.43 & 0.56-0.61 & 0.65-0.72
  \end{bmatrix},
\]
\endminipage\hfill
where
\minipage{0.45\textwidth}
 \[
V_{c}=
  \begin{bmatrix}
    0.66-0.91 & 0.38-0.71 &  0.00-0.32\\
    0.04-0.69 & 0.29-0.80 & 0.57-0.80 \\
    0.24-0.54 & 0.44-0.72 & 0.59-0.76
  \end{bmatrix}.
\]
\endminipage\hfill

 The value thus obtained from above matrix is $39.73^{\circ}-48.47^{\circ}$ having centre value $\theta_{23}^{PMNS}= 44.24^{\circ}$.

 In case of inverted hierarchy

\minipage{0.45\textwidth}
 \[
U_{PMNS}=
  \begin{bmatrix}
 0.67-0.86 & 0.40-0.66 & 0.02-0.36 \\
 0.18-0.57 & 0.37-0.67 & 0.62-0.77 \\
 0.40-0.44 & 0.58-0.63 & 0.61-0.68
  \end{bmatrix},
\]
\endminipage\hfill
where
\minipage{0.45\textwidth}
\[
V_{c}=
  \begin{bmatrix}
    0.67-0.90 & 0.39-0.70 &  0.00-0.33\\
    0.06-0.66 & 0.30-0.75 & 0.63-0.81 \\
    0.31-0.54 & 0.51-0.71 & 0.59-0.71
  \end{bmatrix}.
\]
\endminipage\hfill

The value of $\theta_{23}^{PMNS}$ obtained for inverted hierarchy is $42.70^{\circ}-52.38^{\circ}$ with centre value as $\theta_{23}^{PMNS}= 47.16^{\circ}$.

\begin{figure}[htp]
\centering
\includegraphics[width=.3\textwidth]{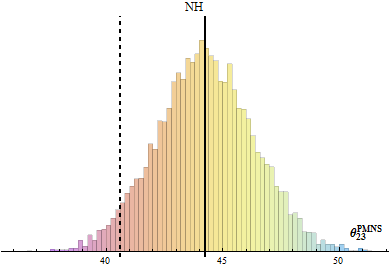}\quad
\includegraphics[width=.3\textwidth]{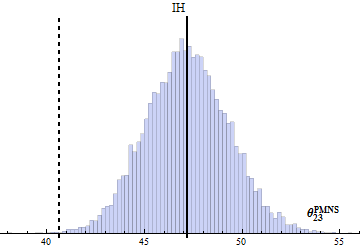}

\medskip

\includegraphics[width=.4\textwidth]{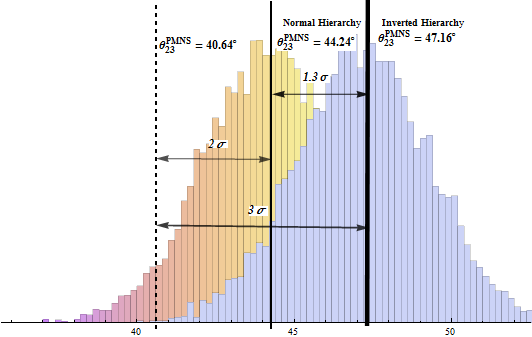}

\caption{Probability density distribution(PDF) plots of $\sin^{2}\theta_{23}^{PMNS}$ including $\lambda$-terms upto $6^{th}$ order for Normal and Inverted neutrino mass hierarchy and their comparison with value obtained in \cite{7}.}
 \label{46theta}
\end{figure}

We have shown the probability density distribution of $\theta_{23}^{PMNS}$ for normal hierarchy (left) and inverted hierarchy (right) and their comparison with the peaked value of $\theta_{23}^{PMNS}$ when hierarchy is not considered\cite{7} in figure~{\bf\ref{46theta}}. Here the dashed line is for $\theta_{23}^{PMNS}$ without considering hierarchy, the thin and thick solid lines are for the two cases of neutrino mass hierarchies i.e. NH and IH, respectively.

\section{Conclusion}\label{sec:concl}
We use the QLC model, where the non trivial relation between the $U_{PMNS}$ and $U_{CKM}$ mixing matrix is taken as the phase mismatch between quark and leptons, via $\psi$ the diagonal matrix. After following the model procedure the central values obtained for $\theta_{23}^{PMNS}$ are $44.24^{\circ}$ and $47.16^{\circ}$ for normal and inverted neutrino mass hierarchies respectively. It has been noticed that the precise values of $\theta_{23}^{PMNS}$ thus obtained for the two cases, NH and IH are about $2\sigma$ and $3\sigma$ away from our previously obtained result\cite{7}, which can give a strong hint for the hierarchy of neutrino masses.

 As such, in future the better precision of $\theta_{23}^{PMNS}$ can give the strong hint about the neutrino mass hierarchy.

\section{Acknowledgments}
B.C. Chauhan and all the other authors acknowledge the financial support provided by University Grants Commission(UGC), Government of India vide Grant No. UGC MRP-MAJOR-PHYS-2013-12281.


\begin{thebibliography}{99}
\bibitem{1} J. Beringer et al (Particle Data Group), Phys. Rev. D28 (2012) 010001; A. Ceccucci, Z. Ligeti and Y. Sakai, PDG (2014) [http://pdg.lbl.gov/2014/reviews/].

\bibitem{2} F. An et al [Daya Bay Collaboration] Phys. Rev. Lett. 108 (2012) 171803; J. Ahn et al [RENO Collaboration] Phys. Rev. Lett. 108 (2012)191802 ; K. Abe et al [T2K Collaboration] Phys. Rev. Lett. 107 (2011) 041801; P. Adamson et al [MINOS Collaboration Phys. Rev. Lett. 107 (2011) 181802; Y. Abe et al [Double Chooz Collaboration] Phys. Rev. D86 (2012) 052008.

\bibitem{3} H. Georgi and C. Jarlskog, Phys. Lett. B 86 (1979) 297 ; H. Minakata and A. Y. Smirnov, Phys. Rev. D 70 (2004) 073009 ; W. Rodejohann, Phys. Rev. D 69 (2004) 033005; K. A. Hochmuth and W. Rodejohann, Phys. Rev. D 75 (2007) 073001; S. K. Agarwalla, M. K. Parida, R. N. Mohapatra and G. Rajasekaran, Phys. Rev. D 75 (2007) 033007.

\bibitem{4} Z. z. Xing, Phys. Lett. B 618 (2005) 141; A. Dighe, S. Goswami and P. Roy, Phys. Rev. D 73 (2006) 071301; A. Y. Smirnov, arXiv:hep-ph/0604213; P. H. Frampton, S. T. Petcov and W. Rodejohann, Nucl. Phys. B 687 (2004) 31; A. Datta, L. Everett and P. Ramond, Phys. Lett. B 620 (2005) 42; J. Harada, Europhys. Lett. 75 (2006) 248; ; J. Harada, Europhys. Lett. 103, (2013) 21001; M. Picariello, B. C. Chauhan, J. Pulido and E. Torrente-Lujan, Mod. Phys. Lett. A 22 (2008) 5860.

\bibitem{5}  B. C. Chauhan, M. Picariello, J. Pulido and E. Torrente-Lujan,Eur. Phys. J. C50 (2007) 573.

\bibitem{6}  Ross, G.G. (1985). Grand unified theories. Westview Press, Reading, MA.

\bibitem{king}{Bj\"{o}rkeroth, Fredrik {\it et al}, Testing constrained sequential dominance models of neutrinos, arXiv:1412.6996 [hep-ph] (2015)}.


\bibitem{7} Gazal Sharma and B. C. Chauhan, JHEP 1607 (2016) 075.



\end{thebibliography}
\end{document}